# Multi-layer architecture for efficient steganalysis of UnderMp3Cover in multi-encoder scenario

Hamzeh Ghasemzadeh

*Abstract*—Mp3 is a very popular audio format and hence it can be a good host for carrying hidden messages. Therefore, different steganagraphy methods have been proposed for mp3 hosts. But, current literature has only focused on steganalysis of mp3stego. In this paper we mention some of the limitations of mp3stego and argue that UnderMp3Cover (Ump3c) does not have those limitations. Ump3c makes subtle changes only to the global gain of bitstream and keeps the rest of bitstream intact. Therefore, its detection is much harder than mp3stego. To address this, joint distributions between global gain and other fields of mp3 bit stream are used. The changes are detected by measuring the mutual information from those joint distributions. Furthermore, we show that different mp3 encoders have dissimilar performances. Consequently, a novel multi-layer architecture for steganalysis of Ump3c is proposed. In this manner, the first layer detects the encoder and the second layer performs the steganalysis job. One of advantages of this architecture is that feature extraction and feature selection can be optimized for each encoder separately. We show this multi-layer architecture outperforms the conventional single-layer methods. Comparing results of the proposed method with other works shows an improvement of 20.4% in the accuracy of steganalysis.

## I. INTRODUCTION

SECRET communication has always been of interest to both individuals and states. This need has been addressed differently throughout the history. Existing solutions can be classified into cryptography and steganography. Cryptography turns the message into unintelligible data whereas steganography conceals the existence of the message by hiding it inside another data. There are some shortcomings associated with cryptography alone. For example, it cannot prevent traffic analysis [1] and important information regarding pattern of communications, its duration, and the intended recipient can be acquired. Also, the mere exchange of encrypted message could raise a red flag and make the adversary aware of importance of communications contents. Steganography can address these by adding an extra layer of protection. Steganalysis is the opposite side of steganography and it tries to break steganography and to detect the presence of hidden messages. Additionally, it provides insight into weakness of steganography methods and helps to develop better steganography methods [2]. Audios are common multimedia signals and hence they are good candidates for steganography. In fact, different audio steganography methods have been proposed [3]. Also, steganalysis community has presented effective analysis for them. Audio methods can be divided into non-compressed and compressed domain. Both of these methods were reviewed recently and comparative studies between them were conducted in [4].

First, audio quality metrics (AQM) were used for differentiating between covers and stegos [5]. Another work argued that AQM are not suitable for steganalysis and instead used Hausdorff distance [6]. Work of [7] showed that features extracted from second order derivative of signals are more significant. Ghasemzadeh et al. argued that by definition, human auditory system (HAS) should be insensitive to steganography noise and therefore models of HAS are not suitable for steganalysis [8]. They proposed a new scale that had the inverse frequency resolution of HAS for feature extraction. Finally, effects of different data hiding algorithms on different bit-planes were investigated in [9]. Based on that observation a universal stego-based calibration method was proposed.

In the compressed domain, most of the works have been devoted to only steganalysis of mp3stego algorithm [4]. Mp3stego hides the message during the compression itself [10]. Westfeld showed that the variances of block sizes of stego and cover are different [11]. Later, an ultra-lightweight system for real-time applications such as integrating steganalysis into intrusion detection systems was proposed [12]. Yan et al. noticed that bit reservoirs in stegos and covers have different characteristics [13]. Work of [14] argued that taking the difference between quantization step of consecutive granules improves the significance of steganalysis features. Another

This paragraph of the first footnote will contain the date on which you submitted your paper for review.

H. Ghasemzadeh is with Departments of "Communicative Sciences and Disorders" and "Computational Mathematics Science and Engineering", Michigan State University, Michigan, USA (e-mail: ghasemza@msu.edu).

This paper has supplementary downloadable material available at http://dx.doi.org/10.21227/H2SX0S, provided by the author. The material

includes: (1) Matlab codes simulating the embedding and extraction procedure of the described Ump3c method, (2) an executable file and its corresponding Matlab wrapper for decoding mp3 frames and importing them into Matlab, (3) implementation of the proposed feature extraction. Contact ghasemza@msu.edu for further questions about this work work showed that the quantization step is a band limited signal and therefore low pass filtering could be used for its calibration [15]. Additionally, it was shown that different mp3 encoders have dissimilar behaviors which would affect results of steganalysis. Work of [15] addressed this, by augmenting the steganalysis features with encoder classification features. Detecting traces of data hiding in the modified cosine transform (MDCT) is another possible approach. Generalized Gaussian distribution and statistical measurements of second order derivative of MDCTs were used in [16]. Finally, difference of absolute values of MDCTs across different channels were exploited for steganalysis [17].

Mp3stego has caused a lot of attention from steganalysis community but it has two major shortcomings. First, its cover should be in non-compressed format. Regarding the problem of using de-compressed signal for cover [18], applicability of mp3stego is only limited to signals that have never been compressed before. Second, mp3stego is designed on top of 8Hz encoder and this information helps steganalysis significantly [15]. For example, if we know an mp3 file has not been encoded with 8Hz, we could conclude that it has not been embedded with mp3stego. On the other hand, Ump3c works directly with mp3 format and therefore, it does not have any of these problems. Unfortunately, steganalysis of Ump3c has not been investigated properly. One of the reasons is that the existing implementation is written for Linux and running it on windows could be challenging. To the best of our knowledge, the only work on steganalysis of Ump3c is an adaptation of regular-singular (RS) method [19]. In the RS method samples of signal are grouped and then the effect of applying a flipping function on the noise of each group is used for steganalysis [20].

There are some important aspects of Ump3c that existing paper has not addressed. For example, RS steganalysis was proposed for image and it may not be efficient for audio signals. Also, there are some issues with work of [19] which may affect it generalization. First, only 200 files were used for its evaluation which limits its external validity. Second, embedded messages were only text files. It is known that distribution of bits 1 and 0 are not the same in text files. Using such messages could significantly change distribution of the signal and could bias result of steganalysis. Third, the work did not consider the existence of different encoders which could play a major role in the performance of the system.

This paper tries to fill those gaps. First, Ump3c is analyzed and two problems with its current implementation are discussed. Second, a new set of features based on mutual information (MI) between global gain (GG) and other fields of mp3 is proposed. Third, multiple re-embedding calibration is proposed for improving potency of steganalysis features. Fourth, the effect of different encoders on distribution of GG is investigated and it is shown that performance of steganalysis varies between different encoders. Finally, a novel multi-layer architecture is proposed to account for different encoders. This system has the unique advantage that features extraction and features selection can be optimized separately for each encoder. Through different simulations we show that this novel architecture outperforms the conventional single layer system.

The rest of this paper is organized as follows. Section II introduces different fields of mp3 bit stream and presents analysis of Ump3c. Section III is devoted to the proposed method and its analysis. Simulation results are shown in section IV. After discussing the proposed method in section V, the paper concludes in section VI.

## II. Mp3 bit stream and Ump3c

### A. Mp3 bit stream and it analysis

Mp3 algorithm is among the most popular audio formats and it provides high quality sound for a compression rate of 90%. Mp3 achieves this through a combination of different techniques including perceptual coding, non-uniform quantization, and Huffman coding. The compression algorithm works as follows [21]. Signal is framed into chunks of 1152 samples. After transforming each frame into frequency domain, its psycho-acoustic model is constructed. Later, this model is used to determine inaudible parts of the signal and how much quantization noise is tolerable in each frequency region. Separately, each frame is divided in two equal granules with 576 samples. Then, each granule is transformed into frequency domain through polyphase filter bank and MDCT operation. Now, MDCT coefficients are quantized in a nested loop. In this fashion, the inner loop adjusts the value of quantization step and makes sure that the existing bit budget is enough for storing the quantized MDCTs. On the other hand, the outer loop compares the quantization noise with psycho-acoustic model of current frame and makes sure that it is below the masking threshold. Finally, quantized MDCTs with their corresponding side information (SI) are written in the bit stream. Also, SI is the part of mp3 bit stream that stores parameters of encoder for each granule and it is vital for correct decoding of mp3. SI has different fields and the ones that are relevant to this study are described here.

One of the most important part of SI is GG and it stores information regarding the value of quantization step of each granule. Decoder uses this information for dequantization of samples and reconstruction of uncompressed signal.

Typically, audio signals have a wide dynamic range covering both fast transient and smooth frames. Mp3 standard accounts for this phenomenon by defining different block types. Psycho-acoustic model of each frame determines the degree of its stationarity and selects the suitable block type accordingly [21]. In this fashion, the best trade-off between time and frequency domain is achieved. Each granule has a set of fields in SI that determines which block types were used during its encoding.

After quantization, MDCTs are encoded into bits with Huffman tables. Mp3 standard has defined different tables for different frequency regions and there are three table select fields per each granule in SI. Our analysis showed that the first table is more informative, so only the first table select field is used.

Typically, frames of audio signals have different complexities. Therefore, some frames can be reconstructed using fewer bits whereas other may need more. Mp3 standard benefits from a mechanism known as bit reservoir to store the extra bits from simpler frames for later more complex ones. In this fashion, overall bit rate is held constant, but length of each individual frame could vary. Part2_3_length is the field that stores length of each granule.

### B. Analysis of Ump3c data hiding algorithm

Ump3c is a data hiding algorithm that works directly with mp3 bit stream and message embedding is done on the GG of SI [22]. Therefore, it doesn't have limitations associated with mp3stego. It also uses the parameter of bit spacing for selecting which granules should be used for data embedding. For example, if bit spacing is equal to 2, only odd granules are used for data hiding. Maximum capacity of Ump3c is 1bit/granule for bit spacing of 1. There are about 38.28 frames in every second of mp3, so, the maximum capacity of Ump3c is 76.56 bit/s.

Our investigations showed some problems with the current implementation of Ump3c that could bias results of steganalysis. First, the message is not encrypted before embedding, therefore if symbols of message are not distributed uniformly, steganalysis results would artificially be higher. To show this, an experiment was conducted. 100 covers were selected randomly and then they were embedded with three different types of random messages. Probability of zero bit was 25%, 50%, and 75% in the first, second, and third case. Then, distribution of GG for each case was estimated separately. Fig. 1 shows the result.

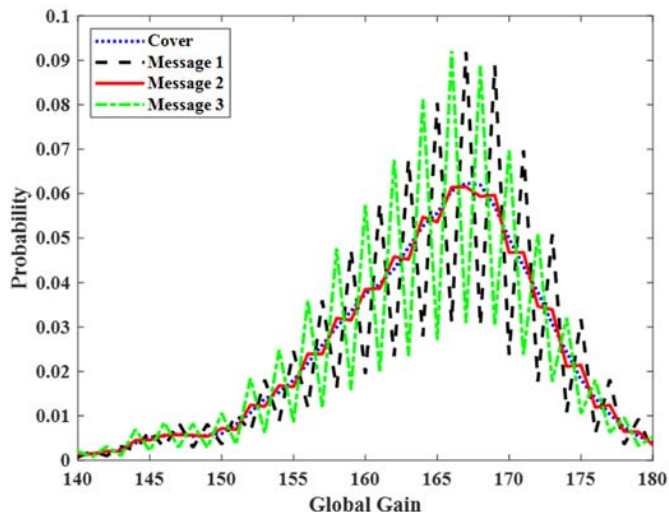

Fig. 1.  Distribution of global gain of covers vs. stegos

Referring to fig. 1 it is evident that distribution of GG between covers and stegos embedded with message type 1 and 3 are quite different. More specifically, probability of odd values is higher (lower) than even values in case 1 (case 3). Also, it is quite evident that in case 2 (which corresponds to uniform random message) distributions of GG of covers and stegos are very similar. It is known that bit stream of encryption systems has uniform distribution, so encrypting messages prior to the embedding could make steganalysis harder and add to the security of Ump3c.

Second, Ump3c select granules consecutively. Therefore, regardless of embedding capacity granules of the first few frames are always embedded with message. Steganalysis system could exploit this faulty implementation and only extract features from specific frames. Recently, it was shown that this characteristic could substantially improve performance of steganalysis [15].

Based on these arguments and to address these problems, a modified version of Ump3c was simulated in Matlab. This new version XOR the message with a pseudo-random sequence and also selects embedding granules randomly. It is noteworthy that XORing the message with a pseudo-random sequence is not totally secure from cryptography point of view, but it is very lightweight, already available in Matlab, and the output sequence has a uniform distribution. This new implementation could also address the difficulty of running Ump3c in windows environment. This new implementation of Ump3c is made publicly available and was used for the rest of simulations.

### III. THE PROPOSED METHOD

#### A. Mutual information

GG is an important parameter of mp3 and when the inner loop increases value of GG many other parts of bit stream change consequently. More specifically, using a larger value of GG decreases the number of bits that are assigned to the granule. Also, it can affect the Huffman tables that are selected, the characteristics of bit reservoir, and even MDCTs. Therefore, embedding a message with mp3stego could change most of the bitstream. On the other hand, in Ump3c only GG field is changed, and all other parts of bit stream are kept intact. Therefore, statistical features from other parts of mp3 do not have any steganalysis relevance and therefore its steganalysis is much harder. Also, referring to fig. 1 we see that for uniform messages, GG of covers and stegos have the same distribution. Thus, simple statistical moments of GG would not be discriminative.

Our solution to this dilemma is as follows. We can think of mp3 as a function that takes an uncompressed signal as the input and produces a multivariate model. In this fashion, the mp3 function imposes certain statistical relationships between its output variables. Based on these terminologies, the true output of mp3 function would satisfy those statistical characteristics. On the other hand, if one variable of this multivariate model is changed, it is possible that some of those statistical relationships would be violated. Therefore, we hypothesized that there would be some deviations between joint probability of GG and other parts of bit stream. We could use those deviations and implement a powerful steganalysis method. But, using the joint probability could be problematic. For example, if 25 bins are used for estimation of joint probability between GG and four different fields of bit stream, the total number of features would be 2500. To tackle this, we have used another approach for capturing deviations in the multivariate statistical model of mp3.

Mutual information (MI) is a useful metric for measuring the amount of information that we can acquire about a random variable ($X$), by observing another random variable ($Z$). Let $p(x,z)$ denote the joint probability of random variables $X$ and $Z$ and $p(x)$ and $p(z)$ denote their marginal distributions. MI between $X$ and $Z$ is denoted by $I(X; Z)$ and for discrete random variables it is defined as:

$$I(X; Z) = \sum_{z \in Z} \sum_{x \in X} p(x, z) . log(\frac{p(x, z)}{p(x).p(z)}) \tag{1}$$

### B. Significance of MI features

Let $x$(m), $y$(m), $n(m)$ and $z(m)$ denote cover, stego, steganography noise, and another signal that has some dependence on $x(m)$. We can write:

$$y(m) = x(m) + n(m) \tag{2}$$

In order to show the potency of MI features we compare $I(X; Z)$ with $I(Y; Z)$. Assuming $n(m)$ is independent from $x$(m) and $z(m)$ and it has zero mean [8, 9, 15, 23-25] we can write:

$$\mu_Y = \mu_X, \qquad \sigma_Y^2 = \sigma_X^2 + \sigma_N^2 \tag{3}$$

where, $\mu$ is the mean and $\sigma^2$ is the variance. Assuming $Y$ and $Z$ have a bivariate normal distribution with correlation coefficient of $\rho_{YZ}$, $I(Y; Z)$ is:

$$I(Y; Z) = -\frac{1}{2} ln(1 - \rho_{YZ}^2) \tag{4}$$

$$\rho_{YZ}^2 = \frac{cov^2(Y,Z)}{\sigma_Y^2 \sigma_Z^2} \tag{5}$$

$$cov(Y, Z) = \mathbb{E}[(Y - \mu_Y)(Z - \mu_Z)] \tag{6}$$

where, $\mathbb{E}$ and $cov$ denote the mathematical expectation and covariance, respectively. Replacing for $Y$,

$$cov(Y, Z) = \mathbb{E}[(X + N - \mu_X)(Z - \mu_Z)] =$$
$$\mathbb{E}[(X - \mu_X)(Z - \mu_Z)] + \mathbb{E}[N(Z - \mu_Z)] \tag{7}$$

$$\mu_N = 0 \Rightarrow \mathbb{E}[N(Z - \mu_Z)] = 0 \tag{8}$$

therefore,

$$cov(Y, Z) = cov(X, Z) \tag{9}$$

On the other hand, based on (3),

$$\sigma_Y^2 > \sigma_X^2 \Rightarrow \frac{cov(Y,Z)^2}{\sigma_Y^2 \sigma_Z^2} < \frac{cov(X,Z)^2}{\sigma_X^2 \sigma_Z^2} \Rightarrow 1 - \rho_{XZ}^2 < 1 - \rho_{YZ}^2 \tag{10}$$

Logarithm is an increasing function, therefore,

$$-ln(1 - \rho_{XZ}^2) > -ln(1 - \rho_{YZ}^2) \Rightarrow I(X; Z) > I(Y; Z) \tag{11}$$

Therefore, MI could reflect the presence of hidden messages.

### C. The proposed features

Based on arguments of previous section we have used MI between GG and other fields of mp3 bit stream to detect trace of hidden messages. Let $G_{i,j}$ denote GG of granule $i$ ($i \in \{1,2\}$) of frame $j$ ($1 \leq j \leq N$) where $N$ is the number of frames in the bit stream. Also, let $g_k$, $b_k$, $p_k$, $t_k$ denote GG, block type, part_2_3_length, and table select of mp3 bit stream that are arranged consecutively ($1 \leq k \leq 2N$). For example,

$$g_k = [G_{1,1}, G_{2,1}, G_{1,2}, G_{2,2}, \dots, G_{1,N}, G_{2,N}] \tag{12}$$

The proposed features can be categorized into three sets. The first set contains 5 features and it includes:

$$\mathbb{F}_1 = [I(G_{1,1:N}; G_{2,1:N}), I(G_{1,2:N}; G_{2,1:N-1}), I(G_{1,1:N-1}; G_{1,2:N}), I(G_{2,1:N-1}; G_{2,2:N}), I(g_{1:2N-1}; g_{2:2N})] \tag{13}$$

The second set captures MI between GG and other fields of mp3 and it includes:

$$\mathbb{F}_2 = [I(g_{1:2N}; b_{1:2N}), I(g_{1:2N}; p_{1:2N}), I(g_{1:2N}; t_{1:2N})] \tag{14}$$

Finally, our analysis showed that tail of distribution of GG in some stegos are different from their covers. To capture those, third set was calculated as maximum and minimum of $G_{1,1:N}$ and $G_{2,1:N}$.

$$\mathbb{F}_3 = [max(G_{1,1:N}), max(G_{2,1:N}), min(G_{1,1:N}), min(G_{2,1:N})] \tag{15}$$

### D. Multiple re-embedding calibration

Finding a set of features that are independent from the content of signal and only reflect the presence/absence of hidden messages is very hard. Steganalysis methods address this through calibration techniques. An ideal calibration method is like an oracle that takes a signal as its input and always produces its corresponding cover (or stego in some cases). Estimating the cover through noise removal [5], down-sampling [26], low-pass filtering [15], and desynchronizing jpeg blocks [27] are some possible implementations of a cover-based oracle. Another possibility is, re-embedding which implements a stego-based oracle [9]. In this paper we extend the idea of re-embedding and propose the multiple re-embedding oracle for feature calibration.

Let $x(t)$, $\mathcal{A}_{em}$, $m(t)$, and $\mathfrak{F}()$ denote a signal, the embedding algorithm that we are trying to detect, a random message, and the feature extraction function. The re-embedded calibrated features were defined as the difference between features of signals $x(t)$ and its re-embedded version when the message $m(t)$ was used [9]. The calibrated features vector $\mathbf{F}$ was defined as:

$$\mathbf{F} = [\mathfrak{F}(x(t)) - \mathfrak{F}(\mathcal{A}_{em}(x(t), m(t)))] \tag{16}$$

Referring to equations 13-15, three different sets of features were defined. Furthermore, because $m(t)$ is a random signal, $\mathbf{F}$ will be a random vector with a certain distribution. Multiple re-embedded calibrated features are defined as the statistical moments of distribution of vector $\mathbf{F}$. In this work we used mean and standard deviation (std) of $\mathbf{F}$. Assuming $\mathfrak{F}$ has $k$ features and it is re-embedded *r*-times, fig. 2 shows the proposed multiple re-embedding calibration procedure.

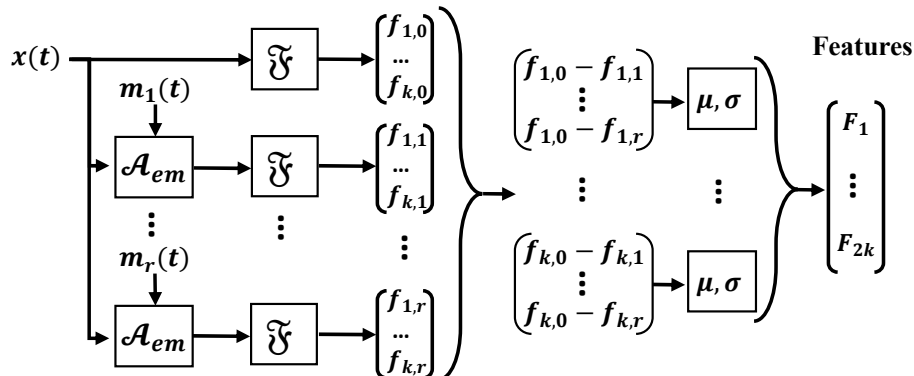

Fig. 2. Multiple re-embedding calibrated features

### E. Effect of different encoders

Previous works have shown that different fields of mp3 have dissimilar distributions [15, 28] and work of [15] showed that almost all aspects of different encoders have quite dissimilar behaviors. That work classified different encoders with only 4 features. Regarding steganalysis of Ump3c, variations between GG of different encoders could play a major role in the performance of steganalysis system. Fig. 3 shows distribution of GG of some encoders.

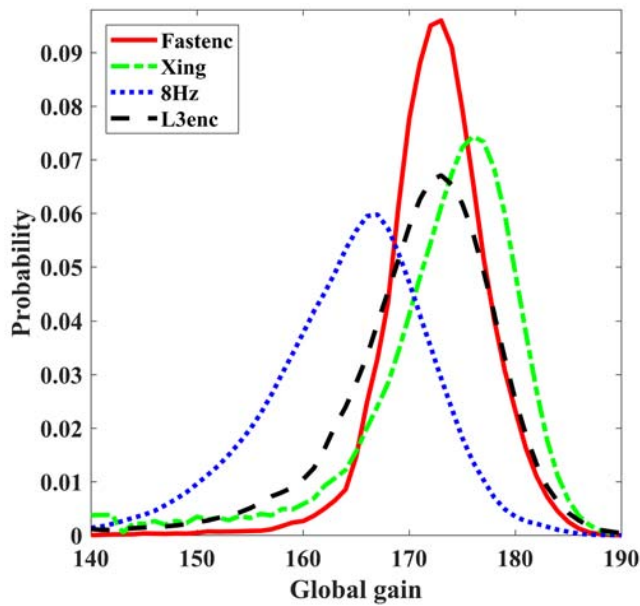

Fig. 3. Distribution of global gain of some encoders

Based on fig. 3 we see different encoders have dissimilar GG distributions. Later in the experiment section we show that if this characteristic is not addressed properly, performance of the system could degrade significantly. One possible solution is to add the encoder classification features to the steganalysis features (single-layer approach) [15]. We argue that for steganalysis of Ump3c this is not a good choice.

First, we need to find $p(x,z)$ before calculation of $I(X; Z)$. Referring to fig. 3 GG has dissimilar distributions for different encoders. Because in the single-layer approach all the files should follow the exact same routine for feature extraction we should find a binning mechanism that is suitable for all encoders. If we use large number of bins we could measure subtle changes in $p(x,z)$ of all encoders but we need more samples to fill the bins reliably. On the other hand, if we use low number of bins fewer samples are needed but we lose precision. Second, due to divergent behaviors of different encoders it is quite possible that different subsets of features are optimum for different encoder. But, in the single-layer approach we should find a sub-set of features that works for all encoders which would be sub-optimum for each individual encoder.

To address these problems, we propose a novel multi-layer architecture where, in the first layer encoders are classified and then in the second layer the actual steganalysis is carried out. The nice thing about this approach is that, steganalysis features could vary for different encoders. Furthermore, feature selection can be done per each encoder and hence the optimum sub-set of features will be used. Fig. 4 shows the proposed multi-layer architecture.

## IV. EXPERIMENTAL RESULTS

### A. Experiment setup

For generating the database, we used 27 audio disks. In this manner we made sure that our excerpts were never compressed before. After splitting all audio tracks into 30 seconds clips, we arrived at 2249 samples. To construct our covers, all samples were then compressed with ten widely used mp3 encoders including 8Hz, Audition, Blade, Fastenc, Gogo, Jetaudio, L3ENC, Lame, Plugger, and Xing. Then, each cover was embedded at 100%, 75%, 50%, 25%, and 12.5% of the maximum embedding capacity with different random messages. Therefore, the final database consisted 134940 files.

For encoder classification we used the 4 features proposed in [15]. Also, in multi-layer method (fig. 4) support vector machine (SVM) in one against one (One-One) arrangement with linear kernel was used for the first layer to achieve the best performance [15]. All other classifications were implemented with binary SVM with radial basis function (RBF) kernel. All tests were carried out with 10-folds cross validation. Also, previous works have shown that normalizing features improves performance of classification [29]. Therefore, for each $l$-th feature, values of mean ($\mu_l$) and std ($\sigma_l$) over all training samples were calculated. Then the normalized features ($\hat{f}_l$) for both training and testing sets were calculated according to (17):

$$\hat{f}_l = (f_l - \mu_l)/\sigma_l \qquad (17)$$

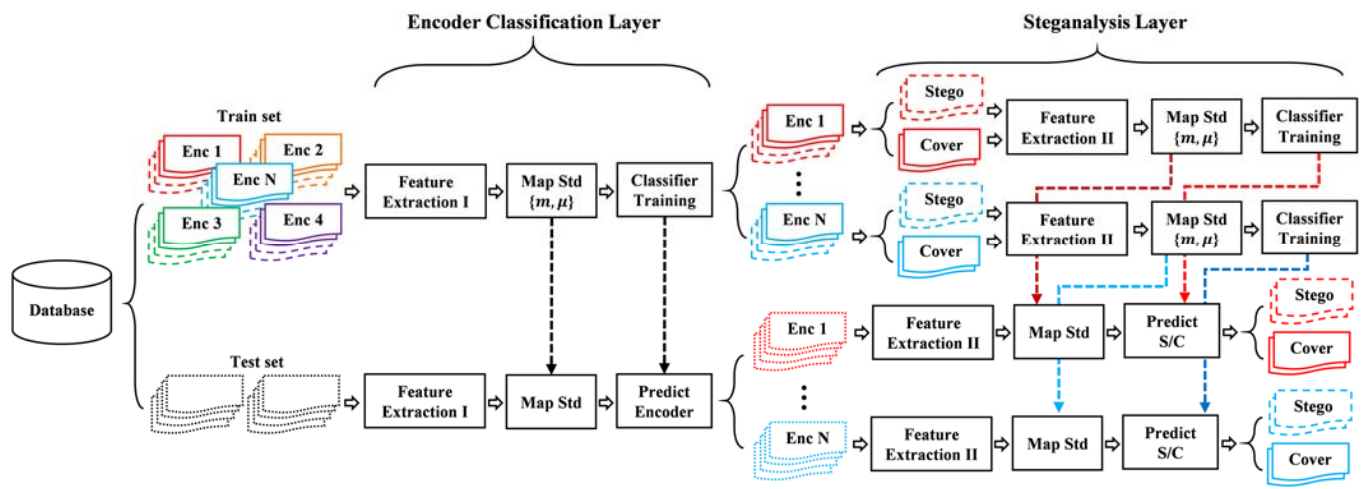

Fig. 4. The proposed multi-layer architecture for steganalysis of Ump3c in multi-encoders scenario

Previous works have compared performance of different feature selections and have shown that genetic algorithm (GA) achieves the best results [29]. Therefore, the best set of features were selected with the help of GA. Details of our GA were as follows. Every generation had 200 individuals and the fitness function was accuracy of classifier. Two point cross-over [30], random replacement, and tournament selection were used for cross-over, mutation, and selection operations, respectively. Finally, in all of simulations steganalysis features are union of sets 1 and 2 (eq. 13, 14), unless otherwise specified.

### B. Optimizing bins of histogram

Estimating the joint probability is the prerequisite of feature sets 1 and 2. Our analysis showed that using different bins for GG, affects significance of those features. Two different cases were compared. In method 1, maximum and minimum of data were extracted and then their distance was divided into 30 equal bins. In the method 2 we had a bin for every possible value of GG. Then, the difference between distributions of GG of covers and stegos for each bin was measured and bins with difference larger than $10^{-5}$ were used for feature extraction. Accuracy of these method are compared in table I.

TABLE I.   PERFORMANCE OF DIFFERENT BINNING STRATEGIES

| Encoders | Method 1 (Embedding Capacity) | | | Method 2 (Embedding Capacity) | | |
|---|---|---|---|---|---|---|
| | 100 | 75 | 50 | 100 | 75 | 50 |
| 8Hz | 87.6 | 81.8 | 73.9 | 90.4 | 85.9 | 78 |
| Audition | 85.9 | 81.3 | 73.1 | 88.4 | 83.9 | 75.5 |
| Blade | 82.9 | 78.8 | 70.4 | 86.8 | 81.6 | 74 |
| Fastenc | 83.3 | 79 | 71.4 | 85.3 | 79.7 | 72.7 |
| Gogo | 81.2 | 76.5 | 68.3 | 85.9 | 80.2 | 73.5 |
| Jetaudio | 62.9 | 59.7 | 56.4 | 63.3 | 60.1 | 56.9 |
| L3ENC | 63.2 | 60.6 | 56.9 | 66.6 | 62.8 | 60.2 |
| Lame | 54.5 | 54.9 | 53.4 | 63.1 | 60.4 | 57.4 |
| Plugger | 95.6 | 93 | 86.1 | 91.8 | 87.5 | 81.4 |
| Xing | 65.5 | 63 | 59 | 67.1 | 62.5 | 59.9 |

Table I shows that improvement as high as 8.6% is achieved when method 2 is used. Therefore, unless otherwise specified, method 2 is used for the rest of simulations.

### C. Optimizing parameters of calibration

The proposed calibration method has two important parameters of re-embedding capacity ($C_R$) and re-embedding order ($r$). A simulation was conducted to measure the effect of different re-embedding capacities on the performance of system. Table II reflects the results.

TABLE II.   EFFECT OF RE-EMBEDDING CAPACITY ($C_R$) ON ACCURACY

| Encoders | Re-embedding capacity | | | | | |
|---|---|---|---|---|---|---|
| | $C_R = 100$ | | $C_R = 75$ | | $C_R = 50$ | |
| | 100 | 75 | 100 | 75 | 100 | 75 |
| 8Hz | 90.4 | 85.9 | 89.4 | 84.6 | 87.9 | 83.3 |
| Audition | 88.4 | 83.9 | 87.2 | 82.5 | 85.1 | 79.9 |
| Blade | 86.8 | 81.6 | 85.2 | 79.4 | 83.3 | 78 |
| Fastenc | 85.3 | 79.7 | 84.6 | 80.1 | 82.9 | 77.5 |
| Gogo | 85.9 | 80.2 | 84.9 | 79.8 | 82.5 | 77.7 |
| Jetaudio | 63.3 | 60.1 | 61.8 | 59.8 | 60.7 | 57.8 |
| L3ENC | 66.6 | 62.8 | 66 | 62.5 | 65.1 | 61.6 |
| Lame | 63.1 | 60.4 | 62.5 | 60 | 60.8 | 58.5 |
| Plugger | 91.8 | 87.5 | 90.2 | 87.2 | 88.1 | 84.8 |
| Xing | 67.1 | 62.5 | 65 | 62.3 | 64.1 | 61.4 |

Calculating mean accuracy of each category leads to 76.7%, 75.7%, and 74.1% for $C_R$ equal to 100%, 75%, and 50%, respectively. We see that re-embedding with capacity of 100% leads to more discriminative features. Therefore, for the rest of simulations $C_R$=100% is used.

Effect of different re-embedding orders was investigated. For a better presentation, average accuracies of all encoders for different embedding capacities and $1 \leq r \leq 10$ are plotted in fig. 5. Based on fig. 5 average accuracy and re-embedding order are directly correlated. But, using higher re-embedding order increases the complexity of feature extraction. Therefore, we used order of 10 for the rest of simulations. Finally, our further investigation showed that the improvement was dissimilar for different encoders. For example, Plugger and L3enc had the highest (7%) and the lowest (1%) improvements, respectively.

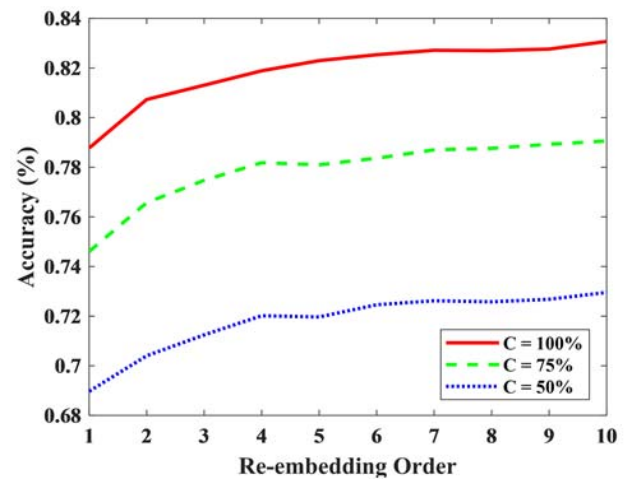

Fig. 5.  Effect of re-embedding orders on average accuracy

### D. Calibrated vs. non-calibrated features

Effectiveness of calibration was investigated. To that end, accuracy of classifier with different feature sets were compared. We tested three different sets. Non-calibrated set which contained feature sets 1 and 2 ($\mathbb{F}_1 \cup \mathbb{F}_2$ from equations 13, 14). Calibrated set which contained calibrated version of feature sets 1 and 2 (when $\mathfrak{F}$s in fig. 2 are replaced with $\mathbb{F}_1 \cup \mathbb{F}_2$). Finally, the extended set was constructed with augmenting calibrated set with feature set 3 (equation 15). Table III reflects accuracy of these three cases. The best results are shown in bold face letters.

TABLE III. ACCURACY OF DIFFERENT FEATURE SETS

|          | Non-Calibrated | | Calibrated | | Extended | |
|----------|------|------|------|------|------|------|
| Encoders | 100  | 75   | 100  | 75   | 100  | 75   |
| 8Hz      | 89.7 | 85.5 | 94.3 | 90.8 | **94.5** | **90.9** |
| Audition | 80.5 | 75.4 | 90.5 | 85.7 | **92.4** | **87.5** |
| Blade    | 86.7 | 82   | **91.8** | **87.5** | 91.7 | 87.2 |
| Fastenc  | 78.1 | 74   | 88.2 | 84.3 | **90.5** | **86.1** |
| Gogo     | 85.8 | 80   | 91.1 | 86.3 | **92.3** | **88**   |
| Jetaudio | 64.9 | 61.4 | 67   | 63.1 | **74.3** | **69.8** |
| L3ENC    | 61.2 | 58.9 | 67.4 | **63.6** | **67.7** | 62.7 |
| Lame     | 57.6 | 55.5 | 65.8 | 61.9 | **93.6** | **88.3** |
| Plugger  | **99.5** | **98.8** | 97.7 | 95.7 | 98.2 | 96.4 |
| Xing     | 68.1 | 63.7 | 69.1 | **66**   | **69.5** | 65.7 |

Referring to table III, we see improvement as large as 10.3% was achieved when calibration was used. Furthermore, when extended set was used, some encoders (ex. Lame) showed significant improvements. Therefore, for the rest of simulations extended feature set was used.

### E. Comparison with previous methods

To the best of our knowledge the only work on steganalysis of Ump3c is RS method [19]. We used the estimated message length of that work as steganalysis feature. Furthermore, both mp3stego and Ump3c change values of GG. Therefore, we analyzed methods published on steganalysis of mp3stego and found that methods of differential quantization step (DQS) [14] and calibrated quantization step (Cal-QS) [15] are also based on statistical analysis of GG. It is noteworthy that in Ump3c, GG is the only part of bit stream that is different between cover and its corresponding stego. Consequently, features extracted from other parts of mp3 bit stream (including MDCT coefficients) have no steganalysis significance. Therefore, methods of [17, 25, 31] were not applicable to steganalysis of Ump3c and were omitted from our comparison. Table IV compares accuracy of the proposed method with those works for different embedding capacities and in the single-encoder scenario. The last row of table shows the average accuracy of all encoders for that specific embedding capacity. The best results are shown in bold face letters.

Table IV shows advantage of the proposed method over existing works. For example, comparing results of Plugger shows that improvement as large as 41.7% is achieved when the proposed method is used. Furthermore, accuracy of all of previous works drops below 60% when embedding capacity is 38.3 bit/s (corresponding to 50% of the maximum embedding capacity) but the proposed method maintains the average accuracy of 64.1% even for embedding capacities as low as 19.1 bit/s. It is noteworthy that at embedding capacity of 19.1 bit/s only $3.88 \times 10^{-3}$% of mp3 bitstream is changed.

### F. Performance in multi-encoder scenario

Previous sections assumed single-encoder scenario. That is, the system was trained and tested for each encoder separately. But, in the real world mp3 files could be generated using different encoders. To test this scenario a new database was generated, where for each file only one of encoders was selected randomly. Three different scenarios for the proposed method were implemented and tested. In the scenario I, extended feature set was used in a single layer architecture. As we discussed in section 3.5 in single layer scenario we cannot optimize binning and feature selection for each encoder. Therefore, we used union of bins from all encoders. Scenario II was basically the first scenario but steganalysis features were augmented with encoder classification features from [15]. Finally, scenario III was implementation of the proposed multi-layer architecture presented in fig. 4. Comparison between accuracy of these scenarios with previous works are presented in table V. The best results are shown in bold face letters.

TABLE V. PERFORMANCE IN MULTI-ENCODER SCENARIO

| Method | Capacity | | | | |
|--------|------|------|------|------|------|
|        | **100** | **75** | **50** | **25** | **12.5** |
| RS          | 61.4 | 58.1 | 55.6 | 52.5 | 50.5 |
| DQS         | 63.7 | 60.8 | 56.8 | 53.6 | 51.1 |
| Cal-QS      | 61.4 | 58.2 | 54.9 | 51.6 | 47.7 |
| Scenario I  | 75.5 | 71.8 | 66.1 | 56.5 | 51.4 |
| Scenario II | 78.6 | 74.1 | 67.1 | 58.7 | 54.7 |
| Scenario III| **83.7** | **79.5** | **73.5** | **61.3** | **55.2** |

Finally, receiver operating characteristic (ROC) of these methods in multi-encoder scenario for embedding capacity of 100% are presented in fig. 6.

TABLE IV. COMPARING ACCURACY OF THE PROPOSED METHOD WITH EXISTING WORKS IN SINGLE ENCODER SCENARIO

|          | Proposed method | | | | | RS | | | | DQS | | | | Cal-QS | | | |
|----------|------|------|------|------|------|------|------|------|------|------|------|------|------|------|------|------|------|
|          | 100  | 75   | 50   | 25   | 12.5 | 100  | 75   | 50   | 25   | 100  | 75   | 50   | 25   | 100  | 75   | 50   | 25   |
| 8Hz      | **94.5** | **90.9** | **83.5** | **69.3** | 59.7 | 67.5 | 64.2 | 60.4 | 56.4 | 63.8 | 60.8 | 57.1 | 53.7 | 69   | 65.4 | 60.2 | 53.7 |
| Audition | **92.4** | **87.5** | **80**   | **66.5** | 53.5 | 65.5 | 60.9 | 58.8 | 54.3 | 63   | 59.9 | 56.6 | 53.5 | 74.6 | 69.5 | 62.9 | 55.1 |
| Blade    | **91.7** | **87.2** | **79.9** | **66.7** | 57.9 | 65.2 | 62.7 | 58.4 | 53.8 | 63.4 | 60.6 | 57.4 | 53.2 | 69.2 | 65.1 | 59.4 | 54.0 |
| Fastenc  | **90.5** | **86.1** | **78.3** | **64.2** | 53.3 | 62.8 | 60.8 | 57.6 | 52.9 | 63.1 | 60.2 | 56.8 | 53.2 | 72.9 | 67.9 | 61.8 | 54.2 |
| Gogo     | **92.3** | **88**   | **81.9** | **67.6** | 58.4 | 67.6 | 63.2 | 60.1 | 54.5 | 75.9 | 70.3 | 64.2 | 57.4 | 65.7 | 61.2 | 57.2 | 51.8 |
| Jetaudio | **74.3** | **69.8** | **61.5** | **53.8** | 49.2 | 51.4 | 51.5 | 49.6 | 49.7 | 58.3 | 56.7 | 54.6 | 52.2 | 68.6 | 63.7 | 57.9 | 51.7 |
| L3ENC    | **67.7** | **62.7** | **60.1** | **52**   | 49.6 | 53   | 53.4 | 51.2 | 50.1 | 54.1 | 53.2 | 51.8 | 50.1 | 52.1 | 50   | 45.7 | 45.2 |
| Lame     | **93.6** | **88.3** | **80.4** | **67.3** | 59.4 | 51.9 | 51.8 | 52.1 | 49.7 | 53.6 | 52.2 | 51.4 | 49.3 | 85.8 | 79   | 69.8 | 61.1 |
| Plugger  | **98.2** | **96.4** | **91.4** | **81.1** | 69.6 | 68.2 | 64.3 | 60.2 | 54.7 | 69.3 | 65   | 60.6 | 55.3 | 74.4 | 69.2 | 62.3 | 55.3 |
| Xing     | **69.9** | **65.7** | **60.4** | **52.4** | 49   | 52.9 | 51.8 | 51   | 50.1 | 53.1 | 52.4 | 52   | 50.0 | 54.1 | 52.2 | 52   | 50.7 |
| Average  | **86.5** | **82.3** | **75.7** | **64.1** | 56   | 66.6 | 58.5 | 55.9 | 52.6 | 61.8 | 59.1 | 56.2 | 52.8 | 68.6 | 64.3 | 59.1 | 53.3 |

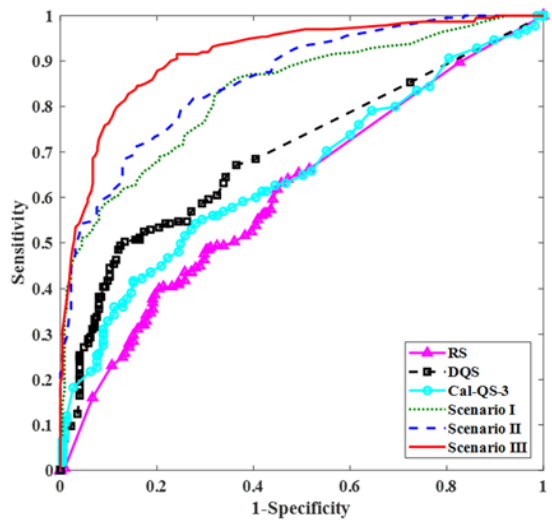

Fig. 6. ROC of different methods in multi-encoder scenario

## V. DISCUSSION

Referring to tables I through IV shows that accuracies of different encoders are quite different. Consequently, we could say that some encoders are more secure for data hiding. This is an interesting observation, because previous works have not investigated the effect of encoders on steganalysis. To track source of these differences we ran a set of analyses on correlation between different fields of mp3 bit stream for different encoders. In one analysis we measured mean of correlation coefficient between GG and its delayed version over all covers for all encoders. Then we sorted different encoders based on two different criteria. In the first case, they were sorted based on average value of correlation coefficient. In the second case, they were sorted based on accuracy of steganalysis for calibrated feature (table III). Comparing the two cases showed comparable orders. Therefore, dissimilar accuracy of different encoders is due to their intrinsic behaviors. That is, encoders with more predictable outcomes are easier to detect and vice versa.

Comparing results of table IV shows that accuracy of the proposed method is much higher than RS which is the best existing method. This shows that different fields of mp3 bit stream are highly dependent on each other. Therefore, if this information is exploited efficiently discriminative property of steganalysis features is improved significantly.

Referring to table V and fig. 6 another strong point of the proposed method becomes evident. Specifically, existing works have very low specificity in multi-encoder scenario. More valuable piece of information is inferred when we compare performance of three different scenarios of table V. Comparing results of scenarios I and II shows that when encoder features are added in single layer architecture performance doesn't improve significantly. On the other hand, when results of scenarios I and III are compared improvements as large as 15.7% is achieved. Therefore, we conclude that the multi-layer architecture is more powerful than single-layer approach. It is noteworthy that application of the proposed multi-layer structure is not limited to this work. For example, better steganalysis system could be constructed based on the proposed architecture. For example, the first layer could classify the content (music vs. speech, different genres of music, different signal complexities, and etc.) and next layer could do the steganalysis part. Also, a universal steganalysis system could be constructed in this way, where each layer uses a different set of features and detects trace of only certain embedding algorithms. Finally, the proposed multi-layer structure is a general framework and is applicable to other classification tasks. That is, the first layer could categorize the data (ex. gender in speech, lighting condition or pose in image, etc.) and next layers do the actual job (ex. speech recognition in speech, face recognition in image, etc.).

Comparing results of table IV with existing works on mp3stego shows a big gap. Specifically, recently it was shown that accuracy as high as 96.6% can be achieved for mp3stego at embedding capacity of 12.5% [15]. Referring to table IV shows that in the similar case (8Hz encoder at capacity of 12.5%) accuracy of Ump3c is only 59.7%. To track source of these differences we ran a set of analyses. In the first analysis we compared the percent of bits that were changed during the embedding process of mp3stego and Ump3c. This result is reflected in table VI. Table VI shows that in mp3stego significant part of bit stream is changed. In the second analysis, distribution of changes in the GG of mp3stego and Ump3c were compared. Our analysis showed that changes in the GG of mp3stego for capacity of 12.5% were in the range of [-43, 21] but for Ump3c with the same embedding capacity the changes in GG were in the range of [-1, +1]. Consequently, the higher accuracy of mp3stego is due to higher distortion that is introduced during the embedding process.

TABLE VI. AVERAGE BITSTREAM MODIFICATION RATE

| Method | Capacity (%) | | | |
|---|---|---|---|---|
| | 100 | 50 | 25 | 12.5 |
| mp3stego | 46.6 | 31.5 | 20.5 | 15.1 |
| Ump3C | 2.96e-2 | 7.54e-3 | 3.88e-3 | 2.0e-3 |

Finally, complexity of the proposed multi-layer structure with single layer structure was compared. For multi-layer structure three different multi-class strategies were tested, one against one SVM (One-One), one against all SVM (One-All), and tree classifier. We used the number of classifiers ($N_C$), training time, testing time, and accuracy at capacity of 100% for this purpose. The results for 10-fold cross validation are reported in table VII.

TABLE VII. COMPARING PERFORMANCE OF DIFFERENT ARCHITECTURES

| Method | $N_C$ | $T_{train}$ | $T_{test}$ | Acc. |
|---|---|---|---|---|
| Single-Layer | 1 | 5.29 | 0.14 | 78.98 |
| Multi-One-One | 55 | 47.3 | 3.62 | 83.59 |
| Multi-One-All | 20 | 144 | 4.69 | 83.12 |
| Multi-Tree | 11 | 2.40 | 1.39 | 83.56 |

Based on results of table VII we see that testing times of all structures are quite acceptable. Therefore, the proposed method does not add any limitations in terms of running time.

## VI. CONCLUSION

Steganalysis of Ump3c in multi-encoder scenario was conducted in this paper. Analyzing embedding mechanism of Ump3c showed that only global gains of bit stream are changed. Therefore, we hypothesized by measuring the joint distribution between global gain and other parts of mp3 bit stream, powerful features could be extracted. To extract relationship between different parts of bit stream and keeping steganalysis features to a minimum, we used mutual information. To further improve significance of the proposed features we introduced multiple re-embedding calibration techniques. Finally, we proposed a novel multi-layer structure for tackling the problem of multi-encoder scenario. This new structure had the important advantage that feature extraction and feature selection can be optimized for each individual encoder. The proposed significantly improved the accuracy of steganalysis in both single encoder and multi-encoder scenarios.

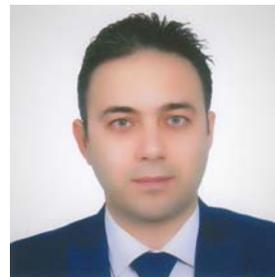

**Hamzeh Ghasemzadeh** was born in Tehran, Iran, in 1984. He received both his BS and MS in Telecommunications Engineering. He was an adjunct professor at department of Electrical Engineering at Azad University, Damavand Branch until 2016. He is now pursuing a dual Ph.D. degree in both "Communicative Sciences and Disorders" and "Computational Mathematics Science and Engineering" at Michigan State University. He has been working on different aspects of audio signal processing, covering security driven applications of audio signals to acoustic analysis of pathological impaired voices. His primary research interests are applying statistical signal processing and machine learning techniques for solving different speech/voice related problems. Right now, he is working on different projects including statistical signal processing and data mining of images acquired through high speed videoendoscopy (HSV) from vocal folds, statistical modeling of connected speech in time and frequency domains, and automatic segmentation of clinical acoustic recordings.